\documentclass[aps,twocolumn]{revtex4}
\usepackage{amsmath,amssymb,graphicx}
\newcommand{\fm}{{\textrm fm}}
\newcommand{\GeV}{\textrm{GeV}}
\newcommand{\MeV}{\textrm{MeV}}
\newcommand{\be}{\begin{eqnarray}}
\newcommand{\ee}{\end{eqnarray}}
\newcommand{\la}{\langle}
\newcommand{\ra}{\rangle}
\begin{document}
\bibliographystyle{apsrev}

\title{Instantons, diquarks and non-leptonic weak decays of hyperons}
\author{M.~Cristoforetti$\mbox{}^1$, P.~Faccioli $\mbox{}^{2\,3}$, 
E.V.~Shuryak $\mbox{}^4$ and M.~Traini $\mbox{}^{3\,5}$}
\affiliation{$\mbox{}^1$ Dipartimento di Fisica, Universit\`a di Milano\\
$\mbox{}^2$ ECT$\mbox{}^\star$, Villazzano (Trento)\\
$\mbox{}^3$ INFN, Gruppo Collegato di Trento\\
$\mbox{}^4$  Department of Physics and Astronomy, 
SUNY at Stony Brook\\
$\mbox{}^5$ Dipartimento di Fisica, Universit\`a di Trento. }
\date{\today}


\begin{abstract}
This work is devoted to the study of  
the non-perturbative contributions in non-leptonic hyperon decays.
We show that the instanton-induced 't~Hooft  
interaction can naturally explain 
the $\Delta~I=1/2$ rule, 
by generating  quark-diquark clustering inside octet baryons.
We compute P-wave and S-wave 
amplitudes in the Instanton Liquid Model, 
and find good agreement with experiment. 
We propose a model-independent procedure 
to test on the lattice if the leading quark-quark attraction 
in the $0^+$ anti-triplet channel responsible for 
diquark structures in hadrons
is originated by the interaction generated by quasi-classical fields 
or it is predominantly due to 
other perturbative and/or confining forces.
\end{abstract}

\maketitle

\section{Introduction}
Weak decays of hadrons encode important
information about the meson and baryon structure and about
the QCD interactions in the perturbative and non-perturbative regimes.
The natural scale of weak processes 
-set by  $W$ boson mass-  is much larger than all other 
scales involved in the hadron internal dynamics. 
This implies that
weak interactions are  effectively local and therefore
can resolve  short distance structures inside hadrons.
Moreover, their explicit dependence on quark flavor and chirality 
can be exploited
to probe the Dirac and flavor structure of the non-perturbative
QCD interaction.

Among the large variety of weak hadronic processes, a prominent role
is played by the 
non-leptonic decays of kaons and hyperons, which are characterized
by the famous $\Delta~I=1/2$ rule \cite{firstdelta12}.
With this name, one refers to the empirical observation that 
amplitudes
in which the total isospin is changed by $1/2$ units
are roughly 20 times larger than the corresponding amplitudes 
 in which the isospin is changed by $3/2$ units. 

Despite nearly 40 years of efforts, the microscopic 
dynamical mechanism responsible for such a striking phenomenon 
is still elusive.
Neither electro-weak nor {\it perturbative} QCD interactions can account for 
the dramatic relative enhancement of the $\Delta~I=~1/2$ decay channels. 
Its origin must therefore
reside in the non-perturbative sector of QCD.

Important insight on the role of non-perturbative dynamics
in non-leptonic hyperon decays  
has come from the observation that in the 
pole-model  (see below) 
the suppression of the decays in the $\Delta~I=3/2$ channel can be 
explained if the quarks participating
to the weak decay are in an anti-symmetric color combination
(Pati-Woo theorem,~\cite{PatiWoo}).
Unfortunately, in a simple Constituent Quark Model 
picture it is not easy to obtain satisfactory
{\it quantitative} predictions for both the P-wave and the S-wave 
amplitudes.
One usually needs to make additional model-assumptions 
on the pole-model part of the amplitude, 
and this somewhat spoils the simplicity of the approach. 
For example, in order to reproduce the data on S-wave amplitudes, 
one needs to include  $1/2^-$ intermediate states~\cite{review}. 

From these considerations it follows that further investigations are still 
needed in order to understand the non-perturbative QCD dynamics underlying
non-leptonic weak decays.
In particular, it would be desirable to set-up a field theoretic
calculation which accounts explicitly for the 
current quark and gluon degrees of freedom.
In this work we explore the possibility that the phenomenology of 
hyperon decays can be understood in the Instanton Liquid Model (ILM).
Such an  approach is derived directly from the QCD Lagrangian, by selecting
a specific set of gauge configurations which are assumed 
dominate the path integral. 

Instantons are topological gauge configurations 
which dominate the QCD path integral in the semi-classical limit. 
They generate an effective quark-quark interaction 
('t~Hooft vertex) which breaks spontaneously chiral symmetry and 
solves the U(1) problem~\cite{thooft1}. 
Evidence for instanton-induced dynamics has been accumulated over 
the years from a variety of phenomenological studies \cite{shuryakrev}
 as well as from 
lattice simulations \cite{cooling,degrand,scalar, pene}. 
In general, these non-perturbative 
vacuum fields  play an important role in the chiral dynamics
 of  light quarks~\cite{dyakonov}, but it is  generally 
believed that they do 
not provide an areal law for the Wilson loop.

The ILM assumes that the QCD vacuum is saturated by an ensemble of instantons
and anti-instantons. The two phenomenological parameters of the model are
the instanton average size ($\bar{\rho}\simeq~1/3~\textrm{fm}$) 
and average density ($\bar{n}\simeq~1~\textrm{fm}^{-4}$). 
These values were first extracted  20 years ago  from 
the global  properties of the QCD vacuum 
(quark and gluon condensates) \cite{shuryak82}.

In the ILM, quarks are bound by the 't~Hooft 
interaction. Even in the absence of confinement, 
the structure of the lowest-lying part of the
light meson and baryon spectra is very 
well reproduced \cite{meson2pt,baryon2pt,mymasses}.
In particular, in this model the lightest octet 
of pseudo-scalar and vector mesons, as
well as the lightest  octet and decuplet of baryons, 
have very realistic masses. 
Moreover, the short-range forces generated by instantons
 allow to reproduce the available 
experimental data on the pion and nucleon
form factors and more  generally  explain the delay of the onset 
of the asymptotic perturbative regime, in  hard exclusive reactions
\cite{pionFF,nucleonFF}.

Besides providing a successful overall description of the light hadron
phenomenology, instantons have a specific property which makes them natural
candidates for the solution of the $\Delta~I~=1/2$ problem.
In fact, Stech, Neubert and Xu pointed out that the body of data
on non-leptonic kaon and hyperon decays can be simultaneously
reproduced, if one assumes that the non-perturbative quark-quark interaction
in the color anti-triplet channel is sufficiently
 attractive to form 
colored quasi-bound structures (diquarks) inside hadrons~\cite{stech}.
Instantons provide a microscopic mechanism which generates 
such a strong attraction binding scalar diquarks \cite{baryon2pt} 
and leading to quark-diquark clustering inside the octet baryons. 

In the past there have been few attempts to understand the $\Delta~I~=1/2$ rule
with instantons \cite{kochelev,inst2}.
In \cite{kochelev} Kochelev and Vento (KV) computed the instanton contribution 
to non-leptonic kaon decays.
On a qualitative level, they found that the 
inclusion of the instanton effects indeed produces
a strong enhancement of the $\Delta~I=1/2$ decay 
channel. On a quantitative level,
such an enhancement was found to be still insufficient 
to reproduce the experimental data.
However, it should be mentioned that non-leptonic
kaon decays in the $\Delta~I=1/2$ channel receive large contribution
also from final-state interactions, which have not been included in 
the KV analysis.
Moreover,  it is now clear that
the  KV calculation is undershooting the instanton 
contribution\footnote{The KV calculation 
was performed in the single-instanton approximation. 
In such an approach, one treats 
explicitly the degrees of freedom of the closest instanton and introduces an
additional parameter $m^*$, which effectively 
encodes contributions from all other instantons.
In their calculation, the authors adopted the phenomenological estimate
for $m^*$ which was available at the time, $m^*\simeq 260$~MeV.
Later, the same parameter was rigorously defined, 
and determined from numerical simulations in the ILM \cite{sia}, 
It was found to be considerably smaller ($m^*\simeq~80$~MeV).}.

In \cite{inst2} the instanton-induced corrections to the effective
Hamiltonian for $\Delta S=1$ transitions were analyzed
in the framework of the
Operator Product Expansion (OPE).
They found that such  ``hard'' instanton effects are rather small.
This result is not surprising: the instanton field cannot
transfer momenta much larger than its inverse size
$1/\bar{\rho}\sim~0.6$~GeV, so instanton effects 
above such a scale are exponentially suppressed.
For this reason, in order to draw conclusions about the role played by 
the 't~Hooft interaction in weak decays, one necessarily 
needs to include their contribution to the ``soft''  hadronic
matrix elements.
In view of these arguments, in the present analysis
we shall neglect all instanton  corrections above the 
hadronic scale set by the inverse instanton size  $\mu=1/\bar{\rho}$
and compute
their contributions to  low-energy matrix elements.

The paper is organized as follows. 
In section \ref{hamiltonian} we analyze the
structure of the effective Hamiltonian for $\Delta~S=1$ transitions 
and explain in detail why instantons are expected 
to produce strong enhancement of the matrix elements associated to 
$\Delta~I=1/2$ transitions. In section \ref{calculation} we 
review the framework which allows to connect parity-conserving
and parity-violating decay amplitudes 
to  low-energy matrix elements of local operators.
The calculation of the decay amplitudes 
 in the ILM is presented in section \ref{calcola}.
In section 
\ref{discussion} we discuss our results and address the question of how
to check  our model assumption of instanton domination
for the light hadron dynamics.
We shall propose  a systematic procedure 
to determine on the lattice if the strong quark-quark 
attractive interaction in the anti-triplet $0^+$ channel
( which drives the $\Delta~I=~1/2$ rule )
is predominantly due to quasi-classical gauge configurations 
or is instead generated by other non quasi-classical fields,
associated to quark confinement.
All results and
conclusions are summarized in section \ref{conclusions}.

\section{$\Delta~S=1$ Effective Hamiltonian and the Origin of the 
$\Delta~I=\frac{1}{2}$ rule} 
\label{hamiltonian}

To lowest order in the Weinberg-Glashow-Salam  Lagrangian, non-leptonic 
weak decays are driven by a single W-boson exchange.
However, such processes receive also QCD and QED
corrections.
These contributions are usually included in the framework of OPE, in which 
one separates short-distance ``hard'' dynamics
from large-distance ``soft'' dynamics.
The former interactions can be treated perturbatively and give rise to 
the well-known effective weak Hamiltonian, which for 
$\Delta~S=1$ transitions reads~\cite{gilman}:
\be
\label{Heff}
H_{eff}^{\Delta S=1}= \frac{G_F}{\sqrt{2}}\,V_{u d} V_{u s}\,
\Big\{\sum_{i=\pm,3,5,6}\,c_i(\mu)\, Q_i + \textrm{h.c.}\Big\}.
\ee
$G_F$ is the Fermi's constant, $V_{u d}$ and $V_{u s}$ are quark mixing 
matrix elements, $Q_i$ are local four-quark operators and $c_i(\mu)$ are
the corresponding Wilson coefficients ($\mu$ is the hadronic scale). 
The local operators $Q_i$ can be written as:
\be
\label{operators}
Q_{\pm}&=&\frac{1}{2}\left[(\bar{u}\,s)_{V-A}(\bar{d}\,u)_{V-A}\,
\pm (\bar{d}\,s)_{V-A}(\bar{u}\,u)_{V-A}\right]
\nonumber\\
Q_{3, 5}&=&(\bar{d}\,s)_{V-A}\,\sum_{q=u,d,s} (\bar{q}\,q)_{V\mp A}\nonumber\\
Q_{6}&=&-2\,\sum_{q=u,d,s} (\bar{q}\,s)_{S+P}\,(\bar{d}\,q)_{S-P},
\ee
where we have adopted the notation 
$(\bar{q}\,q)_{V\pm A}=\bar{q}\,\gamma_\mu~(1\pm\gamma_5)\,q$, and
$(\bar{q}\,q)_{S\pm P}=\bar{q}\,(1\pm\gamma_5)q$.

For a typical hadronic
scale,
$\mu\simeq1$~GeV, the numerical values of the Wilson coefficients are 
$c_+= 0.72$, $c_-=1.97$, $c_3= -0.005$, $c_5=0.003$, 
$c_6=-0.008$~\footnote{For 
an explicit expression of the Wilson coefficients see~\cite{wilsoncoeff}.}.
From these numbers it follows that non-leptonic weak decays are driven by 
the terms proportional to
the operators $Q_+$ and $Q_-$, while all other terms can be neglected.

It is straightforward to verify that the operator $Q_-$ triggers decays
with $\Delta~I~=~1/2$, while the operator $Q_+$ induces transitions
both in the  $\Delta~I~=~1/2$ and in the $\Delta~I~=~3/2$
channel. Hence, in order to explain 
 the $\Delta~I~=~1/2$ rule, one needs to understand the dynamical 
mechanism which enhances the contribution of the term  proportional to $Q_-$.

Accounting only for weak interactions one finds 
$c_+=c_-=1$ and $c_3=c_5=c_6=0$.
Clearly, perturbative strong forces do 
indeed provide a relatively small enhancement  
of  $\Delta~I=1/2$ transitions.
On the other hand, a factor 10 is still missing in order to reproduce
the experimental data. This  must necessarily come 
from the non-perturbative sector of QCD. 
In the OPE formalism, large-distance strong dynamics enters 
through the low-energy matrix elements
of the effective Hamiltonian~(\ref{Heff}). Hence, we conclude that
non-leptonic weak decays are driven by
 non-perturbative forces which enhance by roughly one order 
of magnitude the hadronic matrix elements of $Q_-$,
relative to the matrix elements of $Q_+$.

Significant progress in trying to understand these non-perturbative effects
has been made in a series of works 
by Stech, Neubert,  Xu and Dosch (SNXD)
\cite{stech, stech1,stech2,stech3,stech4}.  
Their starting point was 
the observation that the effective  Hamiltonian 
could be Fierz-transformed into:
\be
\label{crucial}
H_{eff}=\frac{G_F}{\sqrt{2}}\,V_{u d}\,V_{u s} \Big\{
c_-(\mu)\,(u d)^\dagger_{3^*} (s u)_{3^*}  \nonumber\\
 +c_+(\mu)\,(u d)^\dagger_{6} (s u)_{6}  +...+ \textrm{h.c.}\Big\},
\ee
where $(s u)_{3^*}= e_{ikj} s^T_i C (1-\gamma_5) u_j$ is a scalar and
 pseudo-scalar color anti-triplet diquark current, while $(s u)_{6}$ 
is the corresponding color sextet current (the other currents are given by
similar expressions).
From (\ref{crucial}) it follows immediately that the matrix elements 
proportional to $c_-(\mu)$ will be greatly enhanced if the non-perturbative
quark-quark interaction is very attractive in the color anti-triplet channel.
This is most evident in hyperons: if the non-perturbative forces are so 
strong to allow -say- a $s$ and $u$ valence quarks in a $\Sigma^+$
to form a $0^+$ 
anti-triplet quasi-bound state, then these quarks will have a much 
larger chance to be caught in the same point and annihilated by the 
local $(s u)_{3^*}$ operator in the effective Hamiltonian.
A similar argument can be formulated also in the case of kaon
 decays~\footnote{In this case the diquark is formed out of a valence and a 
sea quark.}. 
Based on this simple dynamical assumption, 
SNXD proposed a  phenomenological model which simultaneously 
explains kaon and hyperon non-leptonic decays.

In order to justify the phenomenological assumptions of the SNXD model 
and make contact with QCD, we need to identify some non-perturbative
gauge configurations which, on the one hand, play an important role in the 
hadron internal dynamics and, on the other hand, 
generate color anti-triplet quasi-bound diquarks.
Instantons have precisely this property. In \cite{baryon2pt} it was shown
that the 't~Hooft interaction does indeed form a bound anti-triplet 
\emph{scalar} diquark of mass of roughly 400~MeV.
It is therefore natural to ask whether these fields can provide the microscopic
mechanism underlying the $\Delta~I=~1/2$ rule.

\begin{figure}
\includegraphics[scale=0.3,clip=]{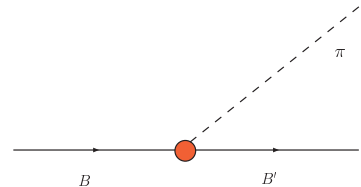} \\
\caption{Factorization
contribution to the non-leptonic hyperon decay 
$\la \pi B'|H_{eff}| B\ra$.}
\label{factfig}
\end{figure}
\section{Low-energy matrix elements}
\label{calculation}

Non-leptonic hyperon decays can be parametrized in terms of two constants 
corresponding to parity-violating and parity-conserving transitions:
\be
\label{PSwave}
\la B'\,\pi| H_{eff} |B\ra = i\,\bar{u}_{B'}\,\left[A-B\gamma_5\right]\,u_B,
\ee
where $B$ ($B'$) denotes the initial  (final)
baryon, and $A$ and $B$ are respectively called
\emph{S-wave} and \emph{P-wave} amplitudes.
The calculation of these amplitudes is generally performed by analyzing
separately two different contributions which correspond to different
mechanisms through which the pion in the final state in
(\ref{PSwave}) can be produced.

In the so-called ``factorization'' part of the amplitude~\cite{BD}, 
the final meson is excited directly
by the color singlet axial-vector current present in the effective 
Hamiltonian (as pictured in  Fig.~\ref{factfig}). 
The corresponding
parity-conserving and parity-violating amplitudes for non-leptonic
hyperon decays with $\pi^-$ in the final state are~\cite{stech}:
\be
A^{\pi^-\,(fact)}_{j\,i} &=&\Big(c_1(\mu)+ 2\,c_6(\mu)\frac{vv'}{m_K^2}\Big)
\,F_{\pi}(M_i-M_j)\,F_{j,i}^{4+i5}\nonumber\\
B^{\pi^-\,(fact)}_{j\,i} &=& -\Big(c_1(\mu)-2\,c_6(\mu)\frac{v^2}{m_K^2}\Big)
\,F_{\pi}(M_i+M_j)\nonumber\\
&\times&\,G_{j,i}^{4+i5}\,
\Big(1+\frac{m_{\pi}^2}{m_K^2-m_{\pi}^2}\Big),\hspace{-2cm}
\label{Bfact}
\ee
with
\begin{eqnarray}
v &=&\frac{m_{\pi}^2}{m_u+m_d}\thickapprox\frac{m_K^2}{m_s+m_u}\nonumber\\
v'&=&\frac{m_K^2}{m_s-m_u}, \qquad F_\pi=~132~\MeV.
\end{eqnarray}
Decay amplitudes with $\pi^0$ in the final state are obtained from the 
substitution
\be
A_{j\,i}^{\pi^0\,(fact)}=-\frac{1}{\sqrt{2}}A_{\pi^-}^{fact}
(c_1\rightarrow-c_2,\ F^{4+i5}\rightarrow F^{6+i7})\nonumber\\
B_{j\, i}^{\pi^0\,(fact)}=-\frac{1}{\sqrt{2}}B_{\pi^-}^{fact}
(c_1\rightarrow-c_2,\ F^{4+i5}\rightarrow F^{6+i7}).\nonumber\\
\ee
In (\ref{Bfact}) 
the $i$ and $j$ indices select the baryons in the initial and final state,
and the constants $F_{j i}$ and $G_{j i}$ are the axial-vector and vector form
factors at zero momentum transfer, defined as:
\begin{eqnarray}
\langle B_j(1/2^+)|J_{\mu}^a|B_i(1/2^+)\rangle_{k_{\mu}\rightarrow 0}
&=& F^a_{j i}\,\overline{u}(j)\,\gamma_{\mu}\,u(i)\nonumber\\
\langle B_j(1/2^+)|J_{5\mu}^a|B_i(1/2^+)\rangle_{k_{\mu}\rightarrow 0}
&=& G^a_{j i}\,\overline{u}(j)\,\gamma_{\mu}\gamma_5\,u(i).\nonumber\\
\end{eqnarray}
Assuming $\textrm{SU}_f(3)$ flavor
 symmetry and using the Goldberger-Treiman relation
we have:
\be 
g_{j i}^a &=&  \sqrt{2}(i\,f_{j a i}\,F + d_{j a i}\,D) g\qquad \textrm{with}
\nonumber\\
g_{j i}^a &=&  \frac{\sqrt{2}}{F_\pi}\,G^a_{j i} \left( M_j + M_i \right)
\nonumber\\
F^a_{j i} &=&  i\, f_{j a i},
\ee
Notice that, in the flavor symmetric limit, the
factorization part of the amplitudes
is completely determined in terms of 
experimentally measured low-energy constants. In this work,
we use the values  \cite{stech2}:
\be
g=13.5,\qquad F+D=1,\qquad \frac{D}{F}\simeq~1.8,
\ee
where the $D/F$ ratio is extracted from semi-leptonic 
decays~\cite{semilept}.
\begin{figure}
\includegraphics[scale=0.3,clip=]{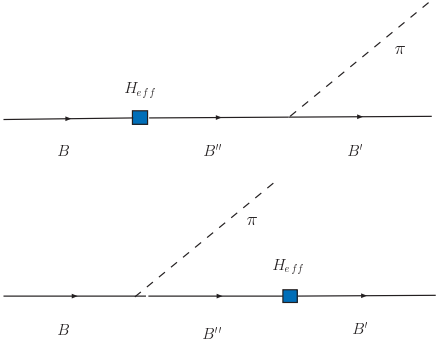} \\
\caption{Pole 
contributions to the non-leptonic hyperon decay $\la \pi B'|H_{eff}| B\ra$.}
\label{polefig}
\end{figure}

It is immediate to verify that factorization amplitudes alone
cannot explain the non-leptonic low-energy decays of kaons and
 hyperons~\footnote{On the other hand, factorization gives
the dominant contribution in energetic B and D non-leptonic
decays~\cite{BD}.}.

The leading contribution to such reactions emerges
from a soft-pion analysis of the matrix element (\ref{PSwave}). 
By applying the PCAC relation, the pion
in the final state is replaced by  an additional operator, expressing
the divergence of the axial-vector current:
\be
\label{PCAC}
\la B_j \, \pi^a(q)| H_{eff}(0)|B_i\ra =\lim_{q^2\to m_\pi^2}
i\,\frac{\sqrt{2}(-q^2+m^2_\pi)}{F_\pi\,m_\pi^2}
\nonumber\\\times \int\,d^4 x\,e^{i\,q x}
\la B_j| T(\partial^\mu\,J^a_{5\,\mu}(x)\,H_{eff}(0))|B_i\ra.\nonumber\\
\ee
One then applies the well-known identity
\be
\label{polesoft}
i\,\int d^4\,x\,T(\,\partial^\mu\,J^a_{5\,\mu}(x)\,H_{eff}(0) )\,e^{i q x}=
\nonumber\\
q^\mu\,\int\,d^4\,x\,T( J_{5\,\mu}^a(x) H_{eff}(0))- i [I^a_5, H_{eff}],
\nonumber\\
\ee
(  $I^a_5$ is the axial charge operator ) 
and performs the analytic continuation
to $q_\mu\to~0$ (soft-pion hypothesis).

The first term in the right-hand side of  (\ref{polesoft}) 
leads to the so-called ``pole contribution''.
Physically, it corresponds to 
the processes in which the effective Hamiltonian
mixes the initial or final baryon  with some intermediate virtual state
(see Fig.~\ref{polefig}).
The final results for the pole contributions read:
\be
\label{polePwave}
B_{j i}^{(pole)} &=& \frac{\sqrt{2}(M_j+M_i)}{F_{\pi}}
\Big[\frac{G_{j l}h^+_{l i}}
{(M_i-M_l)}+\frac{h^+_{j l}G_{l i}}{(M_j-M_l)}\Big]\nonumber\\
A_{j i}^{(pole)} &=& -\frac{\sqrt{2}}{F_{\pi}}
\Big[ E_{j l}\,h_{l i}^- - h^-_{j l} E_{l i} \Big]\nonumber\\
\ee
where $M^*$ denotes the masses of the intermediate
$B_l(1/2^-)$ baryon which is mixed with
the $1/2^+$ baryon by the effective Hamiltonian.
The low-energy constants $h_{j i}^{+(-)}$ and $E_{j i}^a$ are defined as:
\be
\label{hij}
\langle B_j(1/2^+)|H^{pc}_{eff}|B_i(1/2^+)\rangle
&=& h^+_{j i}\,\overline{u}(j)\, u(i)\nonumber\\
\langle B_j(1/2^+)|H^{pv}_{eff}|B_i(1/2^-)\rangle
&=& h^-_{j i}\,\overline{u}(j)\, u(i),\nonumber\\
\langle B_j(1/2^-)|J^a_{5\,\mu}|B_i(1/2^+)\rangle_{k_{\mu}\rightarrow 0}
&=& E^a_{j i}\,\overline{u}(j)\,\gamma_\mu\, u(i),\nonumber\\
\ee

$H_{eff}^{pc}$ is the parity-conserving part of the 
effective Hamiltonian, and reads:
\be
H_{eff}^{pc}= \tilde{A}\,
[ \epsilon_{i j k} (\bar{d}_i C \bar{u}_j)
\epsilon_{l m k} (d_l C \bar{u}_m)+\nonumber\\
\epsilon_{i j k} (\bar{d}_i C\ \gamma_5 \bar{u}_j)
\epsilon_{l m k} (d_l C\ \gamma_5 \bar{u}_m)],
\ee
where
\be
\tilde{A} = \frac{G_F}{\sqrt{2}}\,\sin \theta_c \ \cos \theta_c \, c_-(\mu).
\ee
$H_{eff}^{pv}$ is the parity-violating part of the 
effective Hamiltonian and reads:
\be
H_{eff}^{pv}= -\tilde{A}\,
[ \epsilon_{i j k} (\bar{d}_i C \bar{u}_j)
\epsilon_{l m k} (d_l C\ \gamma_5 \bar{u}_m)+\nonumber\\
\epsilon_{i j k} (\bar{d}_i C\ \gamma_5 \bar{u}_j)
\epsilon_{l m k} (d_l C\ \bar{u}_m)],
\ee

Using $\textrm{SU}_f(3)$ symmetry, one can express  
these matrix elements in terms of few coefficients:
\be
\label{SU3pole}
h^{\pm}_{j i} &=& 2\sqrt{2}\, (i f_{j 6 i}\, f^\pm + d_{j 6 i} d^\pm)
\nonumber\\
 h^-_{j 0} &=& e \delta_{j 6}\nonumber\\
E^{a}_{j i} &=& 2 \sqrt{2}\, (i f_{j a i}\, F^- + d_{j a i} D^-) \nonumber\\
E^a_{0 i} &=& E \delta_{i a}.
\ee

In addition to the pole part, the S-wave amplitudes receive also
a contribution coming from the commutator 
in (\ref{polesoft})~\footnote{In P-wave amplitudes such a contribution 
vanishes because the commutator select only the parity-violating part 
of the effective Hamiltonian.} 
This is usually referred to as  the ``soft-pion'' term: 
\be
\label{Asoft}
A_{j i}^{a\,(soft)}= \frac{-\sqrt{2}}{F_\pi} \, \la B_j |[I^a_5, H_{eff}] |B_i \ra.
\ee

Unlike the factorization part, the pole and soft-pion terms involve
matrix elements which are not directly related to experiments and 
have to be estimated theoretically. 
In the next section we present our calculation of these 
matrix elements in the ILM.

\section{ILM Calculation}
\label{calcola}
In this section we present our calculation
of the  P-wave and S-wave amplitudes, within the ILM.

\subsection{P-wave amplitudes}

In order to determine the P-wave amplitudes in the ILM model, we 
need to evaluate the non-perturbative inputs $h_{j i}^+$, 
defined in (\ref{hij}). 

In a  field-theoretic framework, these
matrix elements  can be extracted
from appropriate ratios of Euclidean three- and two- point functions.
Let us consider the
three-point correlator:
\be
\label{threept}
G^{B' B}_3(\tau) &=&
\int d^3{\bf x}\int d^3 {\bf y}\,\langle 0|\,T(~J^\alpha_{B'}({\bf x},2\,\tau)
\nonumber\\
&\mbox{}&\,\qquad\mathcal{H}_{eff}({\bf y},\tau)
\bar{J}^\alpha_{B}({\bf 0},0)~)|0\rangle,
\ee
where $\tau=i\,t$, $\alpha$ is a spinor index and
$J^\alpha_B(x)$, $J^\alpha_{B'}(x)$ are interpolating operators which 
excite states with the quantum numbers of the  $B$ and $B'$ baryons.
(For example, for the proton and $\Sigma^+$ hyperon we used 
$J^\alpha_{P}(x)= \epsilon_{a b c}\,(u_a^T(x)\,C\,\gamma_5\,d_b(x))\,
u^{\alpha}_c(x)$ 
and $J^\alpha_{\Sigma^+}(x)=\epsilon_{a b c}\,
(s_a^T(x)\,C\,\gamma_5\,u_b(x))\,u^{\alpha}_c(x)$.)

It is straightforward to show that, in the limit of large Euclidean time 
separation,
the correlator (\ref{threept}) relates directly to the matrix element
 $h^+_{B' B}$:
\be
 \lim_{\tau\rightarrow\infty} G^{B'\,B}_3(\tau) = 
2\, h^+_{B' B}\,\Lambda_B'\,\Lambda_{B}
\,e^{-(M_{B'}+M_{B})\,\tau},\nonumber\\
\ee
where $\Lambda_{B'}$ and $\Lambda_{B}$ are the couplings of the 
interpolating fields $J_{B'}$ and $J_{B'}$ to the $B'$ and $B$ states,
defined as
\be
\la 0|J_B(x)|B \ra= \Lambda_B\, u_B(p)\, e^{i\,p\cdot x}.
\ee 
In the $\textrm{SU}_f(3)$ symmetric limit we are considering we have
$\Lambda_{B'}=\Lambda_{B}=\Lambda$ and $M_{B'}=M_{B}=M$.
Hence, in this approximation, it is possible to extract the matrix element
$h^+_{B'\,B}$ by taking the ratio of the three-point 
function (\ref{threept}) with -say- the proton two-point function:
\be
h^+_{B'\,B}=\lim_{\tau\rightarrow\infty}\frac{G_3^{B'\,B}(\tau)}{G_2(2\,\tau)},
\ee
where
\be
\label{twopoint}
G_2(\tau)&=&\int d^3{\bf x}\,
\langle 0|~T[ J^\alpha_P({\bf x},\tau)\bar{J}^\alpha_{P}
({\bf 0},0)]~|0\rangle
\nonumber\\
&\stackrel{\tau\to\infty}{\rightarrow}& 2 \,\Lambda^2 e^{-M\,\tau}.
\nonumber\\
\ee

Non-perturbative calculations of QCD correlation functions  can be 
performed by exploiting the analogy between the Euclidean generating 
functional and the partition function of a statistical 
ensemble.
In lattice QCD, one usually carries-out analytically 
the integral over the fermionic fields, 
and then computes numerically Monte Carlo averages
of the resulting Wick contractions over a statistical ensemble of 
gauge configurations.
In the ILM, we replace the space of all gauge configurations
with an ensemble of instantons and anti-instantons ~\cite{shuryakrev}.
Like in lattice QCD, in each configuration
the quark propagator is obtained
by inverting the Dirac operator. Unlike in lattice QCD, in the
ILM there is no need of regularization, so all calculations are performed in 
the continuum. This prescription
is equivalent to computing
the correlation functions to all orders in the 't~Hooft interaction.

In this work we have considered the simplest version of the 
model, the Random Instanton Liquid (RILM), 
in which the density and size of the pseudo-particles are kept fixed,
while their position in a periodic box and their 
color orientation are generated according to a random
distribution. 

We have evaluated numerically~\cite{cristof} 
the correlation functions associated to the matrix elements
 $\la p|H_{eff}|\Sigma^+\ra$ and $\la \Lambda |H_{eff}| \Xi^0\ra$. 
We have averaged over 52
configurations of 252 pseudo-particles of size $\rho=0.33~\fm$,
in a periodic box of volume $(3.6^3~\times 5.4)~\fm^4$.
Like in lattice simulations, we have chosen a rather large current quark 
mass for $u$ and $d$ quarks (75~MeV), to avoid finite-volume 
artifacts. In order to check for the dependence of our results on the quark
masses, we have also performed the same calculation using  
larger quark masses (135~MeV).
Finally, to enforce flavor symmetry, we have set $m_s=m_u=m_d$.
The 6-dimensional spatial integration in (\ref{threept}) has been 
performed by means of an adaptive Monte Carlo method (VEGAS). Convergence
has been achieved using 1600 integration points. 
The 3-dimensional integral in (\ref{twopoint}) has been
performed by first carrying out the angular integration analytically
(exploiting rotational symmetry) and then computing the remaining 
1-dimensional  radial integration by a Gauss-quadrature method.

We have observed  that the quark-model relation 
$f^+/d^+ \simeq 1$ holds
also in our field-theoretic approach 
\footnote{We remark that, in a field theoretic 
framework, the  relation
$f^+/ d^+\simeq~-1$ is  non-trivial. It is a consequence of the fact that the
sea contribution from fermionically disconnected graphs
is negligible.}, with 
$d^+=~(0.28~\pm~0.05)~\times~10^{-7}~\GeV$, a result
 quite close to the prediction of the SNXD
 model (~$d^+=~0.35~\times~10^{-7}~\GeV$ \cite{stech2}~). 
This calculation shows explicitly that gluon and sea degrees of freedom 
contribute very little to these decay amplitudes.

The results presented so far 
correspond to simulations performed with  quark masses of 75~MeV.
We have found that calculations with heavier quark masses (135~MeV) lead
to very similar results (~$d^+=~0.27~\pm~0.04~\times~10^{-7}~\GeV$ ).
We can therefore conclude that the dependence of these amplitudes 
on the quark mass is very weak.

It is important to ask whether diquark quasi-bound states survive 
within $1/2^+$ baryons, or if they are melted by the 
interaction with the third quark.
To answer, we have compared matrix elements obtained from  
the scalar and from the pseudo-scalar part of the diquark operator 
in (\ref{crucial}).
We have found that such matrix elements are indeed dominated 
by the \emph{scalar} operators in the effective Hamiltonian.
This is a non-trivial result which represents clean signature of the existence
of {\it scalar} diquark structures in the hyperons, in the ILM. 
On the other hand, it also implies that pseudo-scalar diquarks are 
not present in such baryons. 
Finally,  since final-state interaction effects are presumably small in this 
channel~\cite{stech2}, we have neglected them.

Our results for the P-wave amplitudes, 
obtained by collecting the factorization and the pole 
contributions, 
are reported in table \ref{resPwave} and compared to experimental 
data.
\begin{table}
P-Wave Amplitudes~(~$\times~10^{7}$~)\vspace{0.1cm}
\begin{tabular}{rrrrrc}\hline\hline
        \rule{0pt}{3ex} & Pole & Fact.& RILM&Experiment& 
	$\frac{\textrm{RILM}}{\textrm{Exp.}}$  \\ \hline
        \rule{0pt}{3ex}
$\Lambda^0_0$&$-6.87$&$-4.03$&$-10.9\pm1.17$&$-15.61\pm1.4$&$0.7$\\
        $\Lambda^0_-$&$9.72$&$8$&$17.71\pm1.66$&$22.40\pm0.54$&$0.8$\\
        $\Sigma^+_0$&$20.82$&$1.65$&$22.4\pm3.55$&$26.74\pm1.32$&$0.8$\\
        $\Sigma^+_+$&$31.84$&$0$&$31.84\pm4.81$&$41.83\pm0.17$&$0.8$\\
        $\Sigma^-_-$&$1.75$&$-3.26$&$-1.52\pm0.30$&$-1.44\pm0.17$&$1.1$\\
        $\Xi^-_-$&$16.15$&$-2$&$14.15\pm2.75$&$17.45\pm0.58$&$0.8$\\
        $\Xi^0_0$&$-11.42$&$1.01$&$-10.42\pm1.95$&$-12.13\pm0.71$&$0.9$\\
	\hline\hline
\end{tabular}
\caption{Theoretical prediction and experimental results for
P-wave amplitudes. Following the standard notation, 
$B^Q_q$~corresponds to $ \textrm{Amp}(B^Q\to B' + \pi^q)$. 
The RILM prediction is obtained by adding
the pole and factorization contribution. 
 Wilson coefficients have
been evaluated at the hadronic scale $\mu=1/\bar{\rho}=0.6~\GeV$, using
$\Lambda_{\overline{MS}}=230~\MeV$.} 
\label{resPwave}
\end{table}
%
First of all, we observe that the RILM can reproduce the overall body of data
on P-wave hyperon decays. All theoretical amplitudes lie within approximatively
$20\%$ form the experimental results.
Note that 
this discrepancy is of the order of the systematic error introduced by
the assumption of $\textrm{SU}_f(3)$ symmetry.
However, taking a closer look, 
we notice that the central values of the theoretical
predictions consistently undershoot 
the experimental results (except in one case to be discussed below).
This is hardly surprising,
because in the present calculation, 
we have neglected all confining interactions.

Finally, we observe that the theoretical prediction for the
 amplitude $\Sigma^-_-$ is the only one overshooting the
experimental data. 
This is probably a reflection of 
the fact that this is a very delicate channel, where 
the factorization and pole terms are of the same order of magnitude and
have opposite sign.

\subsection{S-wave amplitudes}

S-wave amplitudes, receive contributions from both
the pole and the soft-pion part of the PCAC amplitudes.

The pole part involves mixing of $1/2^+$ baryons with 
$1/2^-$ virtual intermediate states (\ref{hij}). 
As discussed in detail in \cite{stech}, in a simple quark-diquark
model, in order for the $h^-_{j i}$ matrix elements to be non-vanishing
one needs to assume the existence of $0^-$ diquark structures $1/2^-$ 
octet baryons. 
On the other hand, the 't~Hooft interaction 
is repulsive in the $0^-$ channel.
While the attraction in the $0^+$ channel triggers the formation
of scalar diquarks in $1/2^+$ hyperons contributing to P-wave amplitudes,
 the repulsion in the $0^-$ channel
prevents the formation of pseudo-scalar diquarks, which would 
show up in $1/2^-$ hyperons.
Hence, in the ILM, the pole contribution to S-wave amplitudes is expected to 
be suppressed and we shall neglect it.
  
On the other hand, we compute explicitly the soft-pion term (\ref{Asoft}),
which arises from the commutator in (\ref{polesoft}).
For sake of definiteness, let us consider the
$\la P\, \pi^0|H_{eff}|\Sigma^+\ra$ 
S-wave transition. The relevant part of the $Q_-$ operator 
in the effective Hamiltonian can be written in a simplified notation as:
\be
\label{relev0minus}
(d\,u)^\dagger_{0^+}\,(u s )_{0^-} + (d\, u)^\dagger_{0^-} (u s )_{0^+}
+ \textrm{h.c.} 
\ee
The soft-pion contribution depends on the 
commutator of the effective Hamiltonian with the axial-charge operator.
Using current-algebra relationships it is possible to show
that the commutator of (\ref{relev0minus})
with $I_5^a$ gives the same result as
the commutator of the operator:
\be
\label{relev0minus2}
(d\,u)^\dagger_{0^+}\,(u s )_{0^+} + (d\, u)^\dagger_{0^-} (u s )_{0^-}
+ \textrm{h.c.}
\ee
with the $I^a$ operator.
Due to the repulsion of the 't~Hooft interaction
in the $0^-$ diquark channel,  
 the instanton contribution to 
the matrix elements of the second term in (\ref{relev0minus2}) between
$1/2^+$ states is negligible. On the other hand, the matrix elements of 
the first term in (\ref{relev0minus2}) relate to the
 $f^+$ and $d^+$ constants, which have been calculated to determine the
P-wave amplitudes.

Final-state interaction corrections in this channel 
are rather small but not negligible. 
We have included them following the estimate performed
in~\cite{stech1}.

The RILM predictions for S-wave decay amplitudes are presented in
table \ref{resSwave} and compared to experimental results.
As in the case of P-wave transitions, we observe a good agreement 
with experiment, with prediction within
$20\%$ from the data. Again, we observe that the
ILM  tends to undershoot the measured amplitudes, 
which confirms the idea that roughly $20\%$ of the attraction
in the $0^+$ anti-triplet channel comes from confining interactions.
Note that having neglected pole term leads to very reasonable results 
(~except in one channel, $\Sigma^+_+$, where also all other contributions 
vanish).
Clearly, there is no need to assume pseudo-scalar diquark structures in
$1/2^-$ baryons.

\begin{table}
S-Wave Amplitudes (~$\times~10^{7}$~)\vspace{0.1cm}
\begin{tabular}{rrrrrrrr}\hline\hline
        \rule{0pt}{3ex} & soft & fact.& RILM&RILM(FSI)&Experiment& 
$\frac{\textrm{RILM}}{\textrm{Exp.}}$  \\ \hline
        \rule{0pt}{3ex}
$\Lambda^0_0$&$-1.71$&$0.2$&$-1.51$&$-1.75\pm0.34$&$-2.36\pm0.03$&$0.7$\\
$\Lambda^0_-$&$2.41$&$-0.53$&$1.88$&$2.25\pm0.57$&$3.25\pm0.02$&$0.7$\\
$\Sigma^+_0$&$-4.18$&$0.23$&$-3.96$&$-3.55\pm0.64$&$-3.25\pm0.02$&$1.1$\\
        $\Sigma^+_+$&$0$&$0$&$0$&$0$&$0.14\pm0.03$&-\\
        $\Sigma^-_-$&$5.91$&$-0.62$&$5.29$&$4.34\pm0.9$&$4.27\pm0.01$&$1$\\
        $\Xi^-_-$&$-4.83$&$0.61$&$-4.22$&$-4.22\pm0.82$&$-4.49\pm0.02$&$0.9$\\
        $\Xi^0_0$&$3.41$&$-0.22$&$3.20$&$3.20\pm0.58$&$3.43\pm0.06$&$0.9$\\
\hline\hline
\end{tabular}
\caption{
Theoretical prediction and experimental results for
S-wave amplitudes. Following the standard notation, 
$A^Q_q$~corresponds to $ \textrm{Amp}(B^Q\to B' + \pi^q)$. 
The RILM prediction is obtained by adding
the soft-pion and factorization contributions. 
The results in RILM(FSI) include also final-state
interaction corrections, as estimated in \cite{stech2}. 
Wilson coefficients have
been evaluated at the hadronic scale $\mu=1/\bar{\rho}=0.6~\GeV$, using
$\Lambda_{\overline{MS}}=230~\MeV$.} 
\label{resSwave}
\end{table}

\section{Discussion}
\label{discussion}

In the previous section we have shown that the inclusion of instanton-induced
effects allows to reproduce the overall body of data on non-leptonic hyperon
decays.
We recall that both P-wave and S-wave 
results have been obtained by considering only the 
contribution of the operator $Q_-$ in the effective Hamiltonian, which drives
only transitions with violation of isospin 1/2. 
Hence, we conclude that the 't~Hooft interaction does provide
a non-perturbative dynamical explanation of the $\Delta~I~=~1/2$ rule.

An important question to ask is whether
one can rule-out alternative dynamical
mechanisms, which are not based on quasi-classical interactions. 
As already stressed, the essential dynamical property which is required
in order to produce an enhancement of $\Delta~I~=1/2$ 
transitions is an attraction in the scalar anti-triplet $0^+$ channel.
Clearly, any model for the microscopic dynamics which exhibits 
a sufficiently strong attraction in this channel 
will produce scalar diquarks~\footnote{For example, 
also a model built on the extension of the 
one-gluon-exchange interaction into the 
non-perturbative regime will do the job, at least on a  qualitative level.
For a recent quark-model calculation, based on a non-relativistic 
spin-dependent potential see \cite{recent}.}.
It is nevertheless very important to clarify the dynamical origin 
of these structures, whose existence seems to be confirmed by a number of
 independent phenomenological studies 
(for example,  
in connection with  exotic spectroscopy, see~\cite{jaffe,edward}).

In the following we suggest a systematic, model-independent procedure 
to answer the question whether quasi-classical topological
fields do indeed provide
the dominant non-perturbative interactions driving diquark formation and
the  $\Delta~I~=1/2$ rule.
The idea is to evaluate the relevant matrix elements on the lattice
and to compare the behavior under cooling  of the decay amplitudes 
and of the string tension.
\begin{figure}
\includegraphics[scale=0.35,clip=]{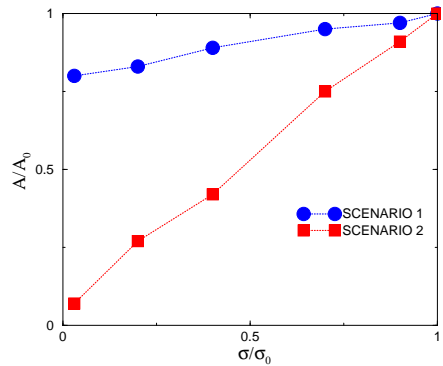} \\
\caption{Two different hypothetical scenarios for the behavior of lattice QCD
decay amplitudes under cooling.
On the x axis, $\sigma/\sigma_0$ represents the
values for string tension obtained after different numbers of cooling steps, 
normalized to the QCD string tension (no cooling). 
On the y-axis, $A/A_0$ represents
the ratio of a decay amplitude computed after the same number of
cooling steps, normalized to its value in QCD (no cooling). 
In SCENARIO~1, the decays are driven
by quasi-classical interactions and the amplitudes change by roughly 
$20\%$ under cooling (ILM prediction). 
In SCENARIO~2, the decays are driven by the confining
forces and vanish rapidly under cooling.}
\label{coolfig}
\end{figure}

The cooling algorithm consists 
of performing statistical averages on different ensembles of
gauge configurations which are closer and closer to the extreme of the 
Euclidean action. This way, the contribution of quasi-classical
fields  is progressively isolated. 
It is well known that, after few cooling steps, all perturbative fluctuations
as well as the confining interactions are removed from the QCD vacuum.
On the other hand, the essential properties of light hadrons,
such as their masses and point-to-point correlators, are seen to 
change very little. This implies that light hadrons are predominantly 
bound by quasi-classical non-confining gauge configurations \cite{cooling}.

The main shortcoming of the cooling procedure is that it leads to 
results which intrinsically depend on the arbitrary number of cooling
steps. Due to this problem, it is  very difficult to make 
systematic, \emph{quantitative} statements.  
On the other hand, the \emph{qualitative} observation that light 
hadrons still
exist in the absence of confinement and that smooth, topological
structures survive
even when the string tension is drastically suppressed are
  model-independent facts, in QCD.

We recall that instantons are smooth, 
topological quasi-classical configurations which bind
hadrons but do not confine.
This observation suggests
to study the behavior of
the  $\Delta~I~=1/2$ decay amplitudes as a function of 
the string tension, calculated after each cooling step 
(see Fig. \ref{coolfig}). On the basis of our analysis we predict that, 
if instantons are indeed the leading dynamical effect, then
the amplitudes should decrease by at most $20\%$, as the
string tension varies from its physical value to nearly zero.
On the other hand, if instantons do not provide 
the dominant interaction in these processes, 
then the amplitudes should drastically
die out, along with the string tension.

$\mbox{}\vspace{1cm}$
\section{Conclusions and outlook}
\label{conclusions}
In this work, we have studied the instanton contribution
to non-leptonic weak decays of hyperons.
We have applied the OPE formalism to separate hard-gluon corrections to soft
non-perturbative effects and we have used the Random Instanton Liquid Model
to compute the relevant low-energy  matrix elements. 
The connection between the matrix elements and the decay amplitudes has been
established considering the contributions arising from 
both the pole and soft-pion terms in the PCAC relations 
and from the factorization part of the amplitude.
Final-state interaction corrections have been applied to S-wave transitions,
and have been neglected in P-wave transitions.

We have found that the ILM yields to a good description of both P-wave and
S-wave decays, providing a microscopic explanation for
the $\Delta~I=1/2$ rule. In this model, the strong enhancement
of the transitions in which the total isospin is changed by 1/2 units
is originated by the strong attraction due to the 't~Hooft interaction 
in the quark-quark scalar anti-triplet channel, leading to a quark-diquark
structure in the hyperons.
We stress that the calculation presented in this work were
performed with no parameter fitting. 
The only phenomenological quantities introduced 
by the ILM are the instanton average size and
density, which have been fixed long ago to reproduce
global vacuum properties.

Our results provide a further confirmation of 
the generally accepted picture according to which 
the internal dynamics of light hadrons is dominated by the interactions
responsible for chiral symmetry breaking.
Indeed in the present calculation, roughly   $70\%$ 
of the amplitudes
comes from instanton-induced interactions (which drive the spontaneous 
breaking of chiral symmetry), 
$10\%$ from hard gluon-exchange corrections, while the remaining  $20\%$ 
is due to some other interactions, presumably related to confinement.
Results are seen to depend very weakly on the value of the current
quark masses chosen.

Since the present analysis is affected by some model-dependence, 
we cannot in principle rule-out possible alternative dynamical mechanisms 
for scalar diquark formation.
However, we have suggested a lattice-based procedure which 
would allow to determine,
in a unambiguous and model-independent way, 
if the strong attraction in the diquark channel is generated  by
 quasi-classical gauge configurations or if it
is due to the quantum fluctuations associated with the dynamics of color 
confinement.

\acknowledgments
We thank D.~Jido and V.~Vento for very useful conversations. P.F. acknowledges 
interesting discussions with J.W.~Negele and R.L.~Jaffe.
Feynmann diagrams have been drawn using JaxoDraw~\cite{jaxodraw}. 

\end{document}